\documentclass{ws-procs975x65}

\begin{document}

\title{PRERAINBOW OSCILLATIONS IN  $^3$He SCATTERING   \\ FROM THE  HOYLE 
STATE OF  $^{12}$C   AND ALPHA PARTICLE CONDENSATION 
}

\author{S. OHKUBO$^*$        }

\address{Department of Applied Science and Environment, \\
Kochi Women's University, Kochi 780-8515, Japan \\
$^*$E-mail: shigeo@cc.kochi-wu.ac.jp\\
www.cc.kochi-wu.ac.jp/\~\  shigeo/}

\author{ Y. HIRABAYASHI }

\address{Information Initiative Center,
Hokkaido University, Sapporo 060-0811, Japan \\
E-mail: hirabay@iic.hokudai.ac.jp}

\begin{abstract}
$^3$He+$^{12}$C scattering is studied in a coupled channel method  
by using a double folding model with microscopic wave
      functions of $^{12}$C.
Experimental angular distributions in elastic and inelastic scattering
 to the   $2^+$ (4.44 MeV),
 0$^+_2$ (7.65 MeV)  and 3$^-$ (9.63 MeV) states of $^{12}$C are well reproduced.
It is found that the Airy minimum of the prerainbow oscillations for the Hoyle
 state is considerably shifted to a larger angle due to its dilute density
 distribution  compared 
 with that of the normal ground state   in agreement with  the idea of 
 $\alpha$ particle  condensation.
\end{abstract}


\bodymatter

 \section{Introduction}\label{intro:sec1}

\par
  Bose-Einstein condensation (BEC) has been well known  in a dilute gas
\cite{Leggett2001}.
The macroscopic  properties such as  superconductivity and 
superfluidity  in both $^3$He  and $^4$He are understood in relation  to BEC.
 Recently it has been  speculated that the 
$0_2^+$(7.65 MeV) state of $^{12}$C,  the Hoyle state, 
  is    a  Bose-Einstein condensate of  three $\alpha$ particles 
  \cite{Tohsaki2001}. 
    Uegaki {\it et al.} 
 \cite{Uegaki1979} and Kamimura and Fukushima  \cite{Kamimura1981}  
studied the $\alpha$ cluster structure of $^{12}$C thoroughly  in the microscopic 
cluster model and showed that the $0_2^+$ state of $^{12}$C,
   has a loosely coupled three $\alpha$ cluster structure with an 
$\alpha$$\otimes$$^8$Be configuration. 
 Recently it has been  shown that the wave functions of Uegaki
 {\it et al.} \cite{Uegaki1979} 
and Kamimura and Fukushima \cite{Kamimura1981}  are  almost completely equivalent 
to the wave function of an $\alpha$ particle condensate  that the three $\alpha$
 particles are sitting in the lowest 0s state
  in  a dilute gas.

A macroscopic property that is peculiar to BEC  such as  superconductivity 
and superfluidity has not been observed in the case of $\alpha$ particle condensation.
Recently Kokalova {\it et al.} \cite{Kokalova2006} proposed a new
experimental  way of testing BEC of $\alpha$
particles in nuclei by directly observing the enhancement of $\alpha$
particle emission  and the multiplicity  partition of the possible
 emitted $\alpha$ particles.  It is important to find a phenomenon which strongly
 reflects the properties of Bose-Einstein condensation of $\alpha$ particles.

We show that  the dilute property of the matter density due to Bose-Einstein condensation
can be seen in the  nuclear  refractive phenomena. 
 Nuclear rainbow scattering has been  powerful in the study of nucleus-nucleus interaction when
 absorption is incomplete \cite{Khoa2007}. 
It has  been  shown that rainbow 
scattering
and the evolution of the Airy minimum can also be  seen in inelastic scattering
 \cite{Michel2004B}.
 It is expected  that the  refractive effect becomes much larger 
and can be seen clearly at  low  incident energies. 
Recently a new concept of prerainbow has been proposed at the lower energy region
 \cite{Michel2002}.
The   refractive index $n$ is  related
 to the optical potential $V$ as follows:
 \begin{equation}
 n(r) = \sqrt{1-\frac{V(r)}{E_{c.m.}}}. 
\end{equation} 
 
There is no useful experimental data in inelastic $\alpha$ particle scattering to the 
Hoyle state 
 in the low  energy region. Fortunately, for $^3$He scattering from $^{12}$C 
there is an experimental angular distribution at 34.7 MeV  measured by 
Fujisawa {\it et al.}  \cite{Fujisawa1973},
 which had been unnoticed
for many years and to which no theoretical attention from the viewpoint of 
 $\alpha$ particle condensation in the Hoyle state had been paid.

\
 \section{ The double folding model}\label{model:sec2}
\par 

\begin{figure}[t]
\begin{center}
\psfig{file=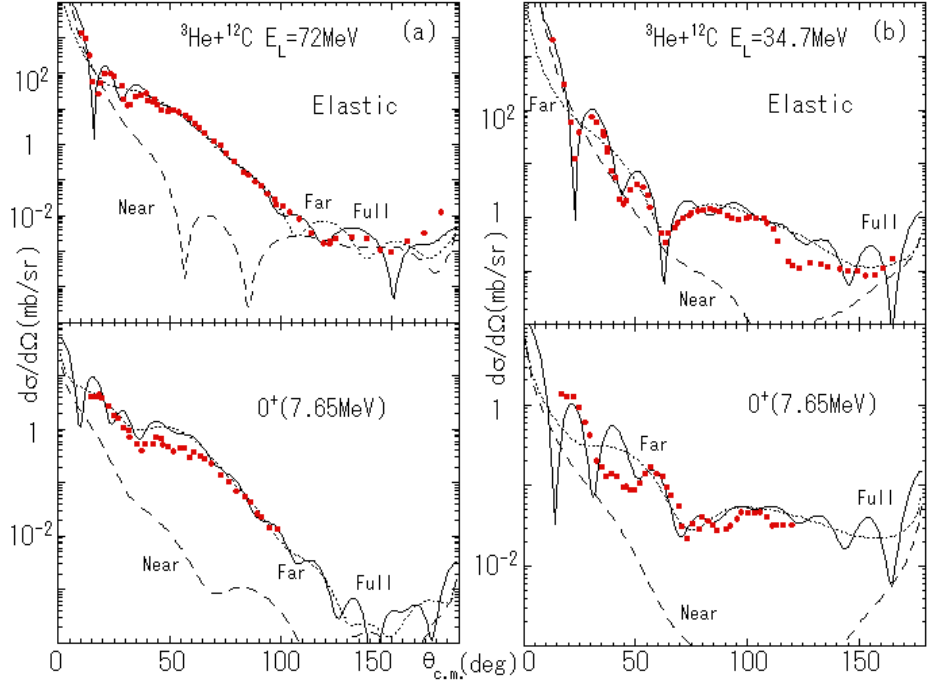,width=5in}
\caption{The calculated angular distributions (solid lines) 
 in $^3$He+$^{12}$C  scattering
 at $E_L$=34.7 and 72 MeV are  decomposed into farside (dotted lines) and nearside 
 (dashed lines)
components and compared  with the experimental data
 (points)\cite{Demyanova1992,Fujisawa1973}.
}
\end{center}
\label{comparison:fig1}
\end{figure}

We study  elastic and inelastic $^3$He+$^{12}$C scattering
  in the microscopic coupled channel method by taking into 
 account simultaneously the  0$^+_1$ (0.0 MeV), $2^+$ (4.44 MeV),
 0$^+_2$ (7.65 MeV),  and 3$^-$ (9.63 MeV) states of $^{12}$C.
The diagonal and coupling potentials for the $^3$He+$^{12}$C
 system are calculated with  the double folding  model:

\begin{equation}
V_{ij}({\bf R}) =
\int \rho_{00}^{\rm (^3He)} ({\bf r}_{1})\;
     \rho_{ij}^{\rm (^{12}C)} ({\bf r}_{2})\;
v_{\rm NN} (E,\rho,{\bf r}_{1} + {\bf R} - {\bf r}_{2})\;
{\rm d}{\bf r}_{1} {\rm d}{\bf r}_{2}  \; ,
\end{equation}

\noindent
where $\rho_{00}^{\rm{ (^3He)}} ({\bf r})$ is the ground 
state density
of  $^3$He taken from Cook {\it et al.}\cite{Cook1981}, while $v_{\rm NN}$ denotes
 the density-dependent M3Y effective interaction (DDM3Y) \cite{Kobos1984}.
$\rho_{ij}^{\rm (^{12}C)} ({\bf r})$ represents the diagonal 
($i=j$) or transition ($i\neq j$) nucleon density of $^{12}$C
calculated in the resonating group method by 
Kamimura {\it et al.} \cite{Kamimura1981}.

The folding potential
is very sensitive to the  wave functions used, which serves
 as a good test of the validity of the 
 wave function. 
  In the analysis  we  introduce the normalization factor 
 $N_R$ for 
 the real part of the potential and phenomenological
 imaginary potentials with a  Wood-Saxon form factor  (volume absorption) 
and a derivative of the  Wood-Saxon form factor (surface absorption) 
  for each channel.

 \section{Analysis of refractive  $^3$He+$^{12}$C scattering}\label{analysis:sec3}
\par
In Fig.~1 angular distributions calculated    using  a coupled channel method  
 at $E_L$=34.7 MeV and    72 MeV
   are compared with the experimental data.
 The calculation reproduces the  experimental  angular
 distributions for the ground state and the Hoyle state
 as well as the $2^+$ 
  and 3$^-$ states  \cite{Ohkubo2007}.

By decomposing the calculated  scattering amplitude into  farside and nearside
 contributions,
the Airy minimum of the rainbow at 72 MeV and the  prerainbow oscillations at 34.7 MeV
 for the Hoyle state is 
 identified. 
At $E_L$=72 MeV the first Airy minimum $A1$  appears at 
 35$^\circ$ for the $0^+_2$ state. For elastic scattering
a clear minimum is not seen in the angular distribution of the 
farside cross sections and the  Airy  minimum
in the experimental data  is obscured  by the interference between
 the farside and nearside amplitudes. 
 On the other hand,
the $A1$ minimum for  the 0$^+_2$ state  is clearly seen in the farside cross sections
 because the minimum is shifted to a larger angle where  the nearside 
  contribution is  much smaller.

 The situation is more clearly seen in the Airy structure at  low the  incident 
 energy region 
where  no typical rainbow 
falloff of the dark side appears.  At $E_L$= 34.7 MeV in  Fig.~1  the Airy minimum $A1$
 appears at 60$^\circ$ for elastic scattering and 75$^\circ$ for the 0$^+_2$ state.
 The latter is much shifted to a larger angle and the Airy minimum is not at all 
obscured 
 by the nearside contributions.
For the 0$^+_2$ state at $E_L$= 34.7 the nearside contributions 
are 
much smaller than the farside contributions compared with the elastic scattering
 case.

Thus the difference of the refraction between the ground state and the 0$^+_2$ state
 is much more clearly seen in the prerainbow oscillations at 34.7 MeV than 
in the rainbow at 72 MeV.
 This shows that the incident 
$^3$He  is strongly refracted in the Hoyle state in  accordance with the previous
 finding  in $\alpha$+$^{12}$C scattering \cite{Ohkubo2004}
that the state has a large lens composed of  three $\alpha$ particles in a dilute
 density distribution.

\begin{figure}[t]
\begin{center}
\psfig{file=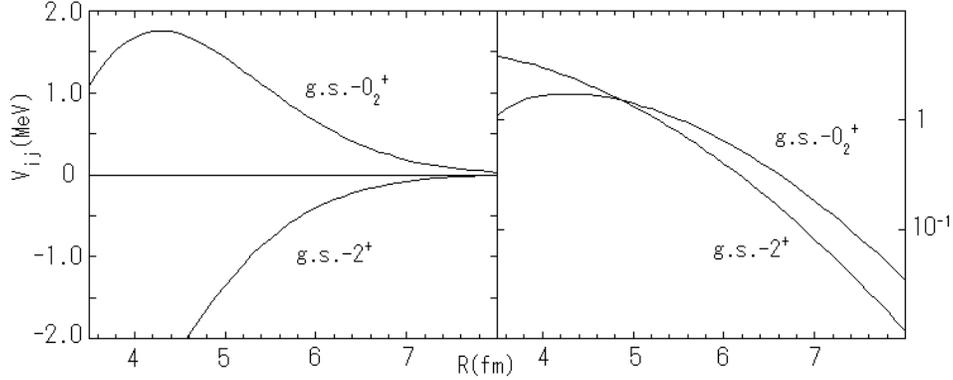,width=5in}
\caption{Coupling potentials for the g.s.-2$^+_1$ and g.s.- $0^+_2$  
channels in $^{3}$He+$^{12}$C scattering at
$E_L$=34.7  MeV 
are shown on a   linear scale (left) and their absolute values are shown on a 
  logarithmic scale (right) to emphasize
 the difference  at the surface region. 
}
\end{center}
\label{coupling:fig2}
\end{figure}

\begin{figure}[t]
\begin{center}
\psfig{file=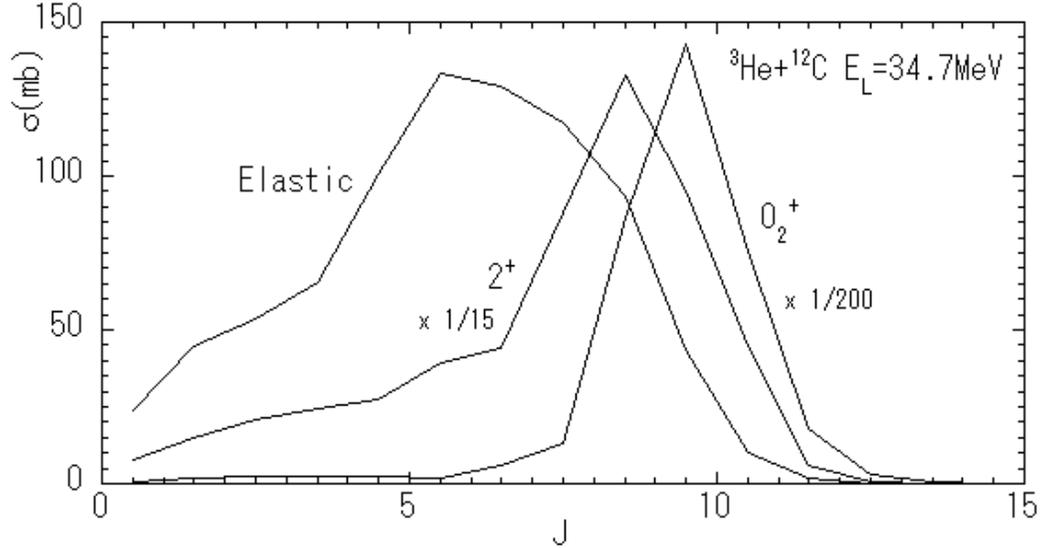,width=5.6in}
\caption{Calculated partial cross sections for elastic  and inelastic 
$^3$He scattering to 
the 2$^+_1$ and   0$^+_2$ state of $^{12}$C at $E_L$=34.7 MeV 
are shown  as a function of total angular momentum.
}
\end{center}
\label{partial:fig3}
\end{figure}

  In Fig.~2   the   coupling potentials for  the g.s.-2$^+_1$ 
and g.s.- $0^+_2$  channels  are compared.  It  is clear
that the coupling potential for the g.s.-$0^+_2$  channel is considerably 
extended to the outer region, which enhances the inelastic scattering 
 to the Hoyle state  
in  the surface region.

 In fact, in  Fig.~3 
we see that  even at the low energy  region where so many high 
 partial waves are not  involved,   inelastic scattering to the Hoyle state occurs
at  large angular momenta (that is, large radius) compared
 with that to  
 the normal g.s. and 2$^+_1$ states. 
This becomes clearer at the high energy region as was discussed 
 in  $\alpha$ particle scattering
from $^{12}$C at $E_L$=139 MeV \cite{Ohkubo2007B}.  
These facts are  in accordance with the previous  finding   that  the
Hoyle state has a large radius compared with the normal ground state.

 \section{Summary  }\label{discussion:sec4}
  We have shown that the prerainbow oscillation is useful for  studying   the dilute density
 distribution due to Bose-Einstein condensation of $\alpha$ particles.  
  The present approach is  applicable  not only to $^3$He scattering but also to 
heavy ion rainbow scattering. The $^{16}$O+$^{12}$C rainbow and prerainbow 
 scattering will be  
useful to reconfirm the present conclusions because the  refractive effect is very strong
 and clear Airy minima of higher order can be expected. 

 It  has been also  suggested that   the  $\frac{1}{2}^-$(8.86 MeV)  state in $^{13}$C 
 , the  $0^+$(9.746 MeV) state in 
$^{14}$C  and the $0^+$( $\sim$29 MeV) state in $^{16}$C  may be a state with 
 one, two and four   
 additional neutrons to the  0$^+_2$ state of  $^{12}$C. 
 If the above states have a dilute  density distribution, the  prerainbow oscillations 
for these  states  would be considerably different from a state with a  normal 
density distribution. 
This kind of experiment 
is highly desired.

 %\section{Summary}\label{summary:sec5}

\section*{Acknowledgments}
One of the authors (S.O.)   has been supported by a
 Grant-in-aid for Scientific Research
 of the Japan Society for Promotion of Science (No. 16540265).

%\begin{verbatim}

%\end{verbatim}
\bibliographystyle{ws-procs975x65}
%\bibliography{ws-pro-sample}

\end{document}